\documentclass{openjournal}

\usepackage{lipsum}

\usepackage{xcolor}
\usepackage{textgreek}
\usepackage[utf8]{inputenc}
\usepackage[english]{babel}

\usepackage{hyperref}
\hypersetup{
    unicode, 
    colorlinks=true,
    linkcolor=linkcolor,
    citecolor=linkcolor,
    filecolor=linkcolor,
    urlcolor=linkcolor,
}
\usepackage{color,colortbl}
\definecolor{linkcolor}{rgb}{0.0,0.3,0.5}
\usepackage{tensind}
\tensordelimiter{?}
\DeclareGraphicsExtensions{.bmp,.png,.jpg,.pdf}
\usepackage{verbatim}
\usepackage[normalem]{ulem}
\usepackage{soul}
\usepackage{xymtex}
\usepackage{newtxtext,newtxmath}
\usepackage{amsmath} 

\def\be{\begin{equation}} 
\def\ee{\end{equation}}

\def\msun{{\msun}}

\def\gsim{\lower.5ex\hbox{\gtsima}} 
\def\lsim{\lower.5ex\hbox{\ltsima}} \def\gtsima{$\; \buildrel > \over 
\sim \;$} \def\ltsima{$\; \buildrel < \over \sim \;$} \def\prosima{$\; 
\buildrel \propto \over \sim \;$} \def\gsim{\lower.5ex\hbox{\gtsima}} 
\def\lsim{\lower.5ex\hbox{\ltsima}} 
\def\simgt{\lower.5ex\hbox{\gtsima}} 
\def\simlt{\lower.5ex\hbox{\ltsima}} 
\def\simpr{\lower.5ex\hbox{\prosima}}   
  
 \def\gtsima{$\; \buildrel > \over \sim \;$} 
\def\ltsima{$\; \buildrel < \over \sim \;$} 
\def\gsim{\lower.5ex\hbox{\gtsima}} 
\def\lsim{\lower.5ex\hbox{\ltsima}} 
\def\simgt{\lower.5ex\hbox{\gtsima}} 
\def\simlt{\lower.5ex\hbox{\ltsima}} 
\def\simpr{\lower.5ex\hbox{\prosima}}

\def\E3{{\cal E}_{\rm g}^{III}} 
\def\msun{\rm M_\odot}

\def\mbh{\rm M_{bh}}
\def\zsun{\rm Z_\odot}

\def\M*{M_*}
\def\mbh{\rm M_{BH}}
\def\mh{M_{h}}
\def\Z*{Z_*}
\def\L*{L_*}

\def\lbol{L_{\rm Bol}}

\def\fej{f_*^{ej}}
\def\feff{f_*^{eff}}

\def\der{{\rm d}}

\def\faccb{f_{\rm bh}^{\rm ac}}
\def\maccb{M_{\rm bh}^{\rm ac}}
\def\med{M_{\rm Edd}}
\def\fed{f_{\rm Edd}}
\def\mcritb{M_{\rm bh}^{\rm crit}}

\def\ergs{\rm erg ~ s^{-1}}

\def \fej {f^{\rm{ej}}_*}
\def \feff {f^{\rm{eff}}_*}
\def \fcold {f_{\rm{cold}}}
\def \fsphinx {f_*^{\rm{sphinx}}}

\begin{document}
\title{Light, heavy, primordial: exploring the diversity of black hole seeding and growth mechanisms in the JWST era}

\author{Pratika Dayal}
\email{pdayal@cita.utoronto.ca}
\affiliation{Canadian Institute for Theoretical Astrophysics, 60 St George St, University of Toronto, Toronto, ON M5S 3H8, Canada}
\affiliation{David A. Dunlap Department of Astronomy and Astrophysics, University of Toronto, 50 St George St, Toronto ON M5S 3H4, Canada}
\affiliation{Department of Physics, 60 St George St, University of Toronto, Toronto, ON M5S 3H8, Canada}

\begin{abstract}
The James Webb Space Telescope (JWST) has revealed a puzzling population of massive black holes in the first billion years, many of which are over-massive compared to their hosts (obese black holes), and reside in metal-poor hosts, posing a challenge for theoretical models at these early epochs. In this work, we compare the observational properties of astrophysically-seeded black holes using the {\sc delphi} semi-analytic model and cosmologically-seeded primordial black holes (PBHs) using the {\sc phanes} analytic model. We explore the growth of light ($\sim 100 \msun$) and heavy ($\sim 10^{3-5}\msun$) seeds through mergers and accretion (both Eddington-limited and at super-Eddington rates) in the astrophysical scenario; PBHs (seeded between $10^{0.5-6}\msun$) only grow through accretion at sub-Eddington rates. Comparing to observables at $z \sim 5-10$, the only model that can be ruled out is the one where we allow Eddington-limited accretion onto light seeds, given their inability of growing to the massive masses observed at these early epochs. The observed high values of the black hole mass-stellar mass relation ($0.3-1$) can be reproduced by both PBHs and heavy seeds accreting at super-Eddington rates. However, only the PBH and Eddington-limited heavy seeding models can simultaneously reproduce the observed black hole masses ($\mbh$), stellar masses ($M_*$), and extremely low host metallicities ($Z \leq 0.01\zsun$) inferred at $z \sim 7-10$. Crucially, we find PBHs show decrease in the black hole mass-stellar mass ratio with increasing halo mass at all redshifts, contrary to any astrophysical black hole model, in addition to inhabiting lower-mass halos. Quantitatively, selecting systems at $z \sim 7$ with $\mbh/M_* > 0.1$ and bolometric luminosities $\sim 10^{44-46} \ergs$ that show a {\it negative} black hole to stellar mass ratio and reside in $10^{9-11}\msun$ halos offer a promising clustering-based discriminant of PBH seeding models.  

\end{abstract}


\maketitle

\section{Introduction}
\label{sec_intro}
The James Webb Space Telescope (JWST) has been revolutionary in shedding light on the emergence of black holes in the first billion years of the Universe. These incredible observations have given rise to a number of puzzles, including: (i) the existence of black holes as massive as $10^8 \msun$ within the first billion years, at redshifts $z \sim 8-10$ \citep{kokorev2023, bogdan2024, kovacs2024}; (ii) black holes that are inferred to be over-massive compared to the stellar masses of their hosts, with black hole to stellar mass ratios at high as $30-100\%$ at $z \sim 5-10$ \citep{ubler2023, kokorev2023, harikane2023bh, kovacs2024, bogdan2024, maiolino2024_jades, kocevski2025,juodzbalis2025, Napolitano2025_Xray}; (iii) black holes hosted in galaxies that are inferred to be metal poor with metallicity values $Z \lsim 0.1-0.01\zsun$ \citep{maiolino2025, tripodi2025}; and (iv) the X-ray weakness of such early black holes \citep{yue2024, Ananna2024, Maiolino2025_Xray, tortosa2026, comastri2026, mazzolari2026}. 

Observationally a number of caveats remain which concern the inferred black hole masses, bolometric luminosities and stellar masses. For example, black hole masses are typically inferred by applying local calibrations to single-epoch measurements of emission lines at high-redshifts. A number of works note that the high velocities indicated by Balmer line broadening could arise due to outflows \citep{baggen2024, matthee2024} and single-epoch local calibrators could over-estimate black hole masses by as much as an order of magnitude \citep{lupi2024,king2024}. On the other hand, dynamical mass measurements yield black hole masses very similar to those inferred from virial estimates, although as of now, these are limited to one lensed black hole at $z=7.04$ \citep{juodzbalis2025b}. Further black hole bolometric luminosities are typically inferred from the H$\alpha$ luminosity after applying dust corrections. This has been a particularly challenging issue for the class of black holes called ``little red dots" that are selected on the basis of their ``V-shaped" spectral energy distribution (blue in the ultraviolet and red in the optical), and compact morphology \citep{kocevski2023, greene2024, setton2025}. Spectroscopy of two such little red dots at redshifts of $z = 3.1$ and $4.46$ have been used to infer bolometric luminosities that are lower by a factor of ten compared to those  from local relations \citep{greene2026}. The stellar mass, too, remains a matter of the debate, especially so in systems that host black holes. Stellar masses are inferred by fitting spectral energy distributions (SEDs) to photometry - assumptions on the initial mass function (IMF), burstiness, and the relative contribution from the black hole and stellar components all play a role in the final inferred stellar mass. For example, changing the slope of the IMF alone is sufficient to change the inferred stellar mass by an order of magnitude \citep[e.g.][]{wang2024}. Further confusing the picture, clustering estimates of 5 faint broad line H$\alpha$ emitters have been used to infer host stellar masses that are about 40 times lower than those inferred from galaxy-only SED fits \citep{matthee2025}. 

With such observational caveats in mind, if one proceeds to take the observations at face value, the prevalence of massive and ``obese" early black holes (with respect to the host stellar masses) pose a crucial challenge to theoretical models. A number of theoretical approaches have been involved to explain such observations which range from: (i) episodic super-Eddington accretion on to light- or heavy-seeds \citep{schneider2023,maiolino2024_jades, volonteri2025, hu2025} although other works seem to rule out super-Eddington accretion as a solution, at least for little red dots \citep{sacchi2025}; (ii) extremely efficient accretion- and merger-driven growth of intermediate-to-heavy seeds \citep{prole2024,caceres2026}; (iii) such systems being the initial phases in the growth of heavy seeds \citep{natarajan2024,pacucci2026}; (iv) selection biases resulting in observations preferentially selecting systems where the black hole outshines the host \citep{volonteri2023,ziparo2026}; (v) baryons being inefficient in star formation \citep{maiolino2024_jades}; or (vi) such objects indicating rapidly spinning black holes \citep{inayoshi2024, caceres2026}. Such astrophysical solutions have recently been supplemented by theoretical models that look to ``cosmological" primordial black holes (PBHs) as a possible solution \citep{dayal2024_PBH, dayal2026_PBH, ziparo2025, prole2025, zhang2026}. A crucial difference between astrophysical seeding models and cosmological primordial black holes is that whilst the former form in halos generated from standard initial density perturbations, the latter can act as the seeds of structure formation, assembling their halos around themselves. Interestingly, primordial black hole models naturally predict extremely low metallicities for black hole hosts, in addition to yielding systems with high black hole-to-stellar mass ratios \citep{dayal2024_PBH,maiolino2025, dayal2026_PBH}. This has made them an attractive alternative to explaining these puzzling early systems.

In this work, we aim at comparing the observational properties of astrophysical and cosmologically seeded black holes. In terms of astrophysically-seeded black holes, we explore light ($100\msun$) and heavy ($10^{3-5}\msun$) seeding models, that can grow both at Eddington and super-Eddington rates. For cosmologically-seeded PBHs, we explore a mass spectrum extending between $10^{0.5-6}\msun$ where we cap accretion at $0.25$ times the Eddington rate, driven by black hole observations at $z \sim 10-10.4$. We start by detailing the models used in Sec. \ref{sec_model}. We discuss the global demographics of early black holes - including the redshift evolution of the black hole mass function and bolometric luminosity function - in Secs. \ref{sec_bhmf} and \ref{sec_blf}, respectively. We then discuss the black hole mass-stellar mass relation in Sec. \ref{sec_mbhms}, the metallicity evolution of black holes hosts in Sec. \ref{sec_metmbh} and the properties of their host halos in Sec. \ref{sec_hosts} before ending with conclusions in Sec. \ref{sec_conc}.

We adopt a $\Lambda$CDM model with dark energy, dark matter and baryonic densities in units of the critical density as $\Omega_{\Lambda}= 0.673$, $\Omega_{m}= 0.315$ and $\Omega_{b}= 0.049$, respectively, a Hubble constant $H_0=100\, h\,{\rm km}\,{\rm s}^{-1}\,{\rm Mpc}^{-1}$ with $h=0.673$, spectral index $n=0.96$ and normalisation $\sigma_{8}=0.81$ \citep[][]{planck2020}. Throughout this work, we use a Salpeter IMF \citep{salpeter1955} between $0.1-100 \msun$. Finally, we quote all quantities in comoving units, and express all magnitudes in the standard AB system \citep{oke-gunn1983}.

\section{The Theoretical model}
\label{sec_model}
In this work, we use two main pathways of black hole seeding and growth: (i) ``astrophysical" - where halos are seeded with either heavy ($10^{3-5}\msun$) or light ($100 \msun$) black hole seeds formed through astrophysical processes. The seeding and growth of such astrophysical black holes is followed using the {\sc delphi} semi-analytic model \citep{dayal2019,dayal2025,caceres2026}; (ii) ``cosmological" - where primordial black holes seeded (by e.g. phase-transitions) at the time of the Big Bang or inflation act as the seeds of structure formation, assembling their halos around themselves. We use the {\sc phanes} analytic model \citep{dayal2024_PBH, dayal2026_PBH} in this case. We briefly describe these models and interested readers are referred to the papers referenced here for complete details. All of the models used here are summarised in Table \ref{table1}.

\subsection{DELPHI semi-analytic model for astrophysical black holes}
\label{delphi}
We use the {\sc Delphi} ({\bf D}ark Matter and the {\bf e}mergence of ga{\bf l}axies in the e{\bf p}oc{\bf h} of re{\bf i}onization) semi-analytic model for galaxy formation to track the seeding and assembly of ``astrophysical" black holes in the first billion years. This model uses a binary merger tree approach to jointly track the build-up of dark matter halos, their baryonic components (gas, stellar, dust and metal masses) and black holes. We follow the assembly of dark matter halos between $\log(M_h/ \msun)=8-14$ from $z \sim 40$ down to $z = 4.5$ in time-steps of 30 Myrs. This time-step is chosen so that all TypeII SN (SNII) from a given stellar population explode within it. Each $z=4$ halo is assigned a co-moving number density by matching to the $\der n / \der M_h$ value of the $z=4$ Sheth-Tormen halo mass function \citep[HMF;][]{sheth-tormen2002} and every progenitor halo is assigned the number density of its $z=4$ parent halo; we have confirmed that the resulting HMFs are compatible with the Sheth-Tormen HMF at all $z \sim 5-20$. 

The very first progenitors (``starting leaves") of any galaxy are assigned an initial gas mass corresponding to the cosmological baryon-to-dark matter ratio such that ${\rm M_{g}^i} = (\Omega_{\rm b}/\Omega_{\rm m}) \mh$. As detailed in \citet{mauerhofer2025}, for each halo, we incorporate interstellar medium (ISM) physics - specially in terms of the fraction of cold gas ($\fcold$) and the stochastic star formation efficiency over the past 30 Myrs ($\fsphinx$) - by using distributions sampled from the {\sc sphinx}$^{20}$ hydrodynamic simulations \footnote{\url{https://sphinx.univ-lyon1.fr/}}. At any time-step, the available gas mass, built-up by both mergers and accretion, can form stars with an ``effective" star formation efficiency $\feff = \min(f_{\rm{cold}} \times f_*^{\rm{sphinx}}, \fej)$ where $\fej$ is the efficiency which produces enough SNII feedback to eject the rest of the gas. The key free parameter in this model is the fraction of SNII energy that can couple to gas ($f_*^w$). We obtain a value of $f_*^w \sim 4\%$ by matching to all available Lyman Break Galaxy data including the redshift-evolution of the ultraviolet luminosity function (UV LF) at $z \sim 5-14$ and the evolving stellar mass function (SMF) at $z \sim 5-10$, as detailed in \citet{mauerhofer2025}.

\begin{table*}
\centering
    \begin{tabular}{|c c c c c c|}
    \hline
     Model & Code & ${\rm M_{seed}} [\msun]$ & $\faccb \lsim \mcritb ~ (\gsim \mcritb)$ & $\fed \lsim \mcritb ~ (\gsim \mcritb)$ & $f_{bh}^w$ \\
    \hline
   {\sc hs-edd1} & {\sc delphi} & $10^{3-5}$ & $5\times10^{-3} (0.1)$ & $10^{-3} (1)$ & $10^{-4}$\\ 
   {\sc hs-edd5} & {\sc delphi} & $10^{3-5}$ & $5\times10^{-3} (0.1)$ & $10^{-3} (5)$ & $10^{-4}$\\ 
   {\sc ls-edd1} & {\sc delphi} & $100$ & $5\times10^{-3} (0.1)$ & $10^{-3} (1)$ & $10^{-4}$\\ 
   {\sc ls-edd5} & {\sc delphi} & $100$ & $5\times10^{-3} (0.1)$ & $10^{-3} (5)$ & $10^{-4}$\\ 
   {PBHs} & {\sc phanes} & $10^{0.5-6}$ & $0.1$ & $0.25$ & $10^{-3}$\\ 
          \hline
    \end{tabular}
    \caption{For the model noted in column 1, we study black holes seeded astrophysically (using the {\sc delphi} semi-analytic model) and cosmologically (using the {\sc phanes} analytic model), as noted in column 2; in the model name, {\sc hs} and {\sc ls} refer to heavy and light seeds, respectively, and the value after {\sc edd} refers to the maximum allowed Eddington accretion rate. Column 3 shows the associated seed mass. Columns 4 and 5 show the fraction of available gas mass and Eddington fraction allowed for accretion onto a black hole in a halo that is less (more) massive than the ``critical" halo mass for efficient accretion noted in Sec. \ref{delphi}. Finally, column 6 shows the fraction of black hole energy that couples to the gas.}
    \label{table1}
\end{table*}

We seed black holes in starting leaves down to $z \sim 13$ \citep{dayal2019}. In this work we explore two seeding models: (i) Light seeds ($\sim 100 \msun$), which are the end products of metal-free population III stars \citep[PopIII; e.g.][]{abel2002, bromm2002}; (ii) heavy seeds ($\sim 10^{3-5}\msun$) which can form in dense, massive stellar clusters through pathways including dynamical interactions \citep[e.g.][]{devecchi2009}, the runaway merger of stellar mass black holes \citep[e.g][]{belczynski2002} or the growth of stellar mass black holes in conjunction with mergers \citep[e.g.][]{leigh2013}; interested readers are referred to Sec. 2.3.1 in the review by \citet{amaro-seoane2023} for more details. These black holes can grow both through accretion and mergers. In terms of accretion, we include a ``critical" halo mass for efficient black hole accretion with a value that evolves with redshift as $\mcritb(z) = 10^{11.25}[\Omega_m(1+z)^3 +\Omega_\Lambda ]^{-0.125}$ (with a scatter of 0.5 dex), motivated by the results of cosmological simulations \citep[e.g.][]{bower2017}. Black holes are allowed a gas accretion rate of $\maccb(z) = min[\epsilon_r \faccb  M_{\rm g}^{\rm sf}, \fed \med]$ where $\epsilon_r=0.057$ is the radiative efficiency for a non-spinning black hole, $\faccb$ is the fraction of available gas that the black hole can accrete, $M_{\rm g}^{\rm sf}$ is the gas mass left after star formation and its associated SNII feedback, $\fed$ is the Eddington fraction and $\med$ is the Eddington accretion rate. Allowing very weak AGN feedback ($0.01\%$ of black hole feedback coupling to the gas) we find values of $\faccb = 0.1 ~ (5 \times 10^{-3})$ and $\fed = 1.0 ~ (10^{-3})$ for halos above (below) the critical mass (we allow 0.5 dex of scatter on all of these quantities) to match observables including the black hole bolometric luminosity function and mass function at $z \sim 5-7$ in the {\it fiducial} model ({\sc hs-edd1}) where heavy seeds can grow capped at the Eddington rate; we refer interested readers to Sec. 3.1 of \citet{caceres2026} for the calibration of free parameters in the model. That is, black holes in halos 
more (less) massive than the critical halo mass can accrete the minimum between 10\% ($0.5\%$) of the available gas mass and $100\% ~(0.1\%)$ of the Eddington fraction in the Eddington-limited scenario; black holes are allowed to accrete at 5 times of the Eddington rate in the super-Eddington limit. Black holes can also grow through mergers in the model - we account for the fact that galaxies and their black holes merge after a ``merging" timescale which can be calculated as \citep{lacey-cole1993}
\begin{equation}
\tau = f_{\rm df} \Theta_{\rm orbit} \tau_{\rm dyn} \frac{M_{\rm host}}{M_{\rm sat}} \frac{0.3722}{ln(M_{\rm host}/M_{\rm sat} )},
\end{equation}
where $M_{\rm host}$ is the mass of the host including all the satellites, $M_{\rm sat}$ is the mass of the merging satellite, $\tau_{\rm dyn}$ represents the  dynamical timescale and $f_{\rm df}=1$ represents the efficiency of tidal stripping. $\Theta_{\rm orbit}$ is well modelled by a log-normal distribution such that ${\rm log} (\Theta_{\rm orbit}) = -0.14 \pm 0.26$ \citep{cole2000}; we randomly sample values from this distribution for each merger. We also make the limiting assumption that satellite galaxies, waiting to merge, neither form stars nor have any accretion of gas onto the black hole. We use the same free parameter values for all of the astrophysical black hole models considered here since our aim is to highlight differences between them. 

\begin{figure*}
\begin{center}
\center{\includegraphics[scale=0.12]{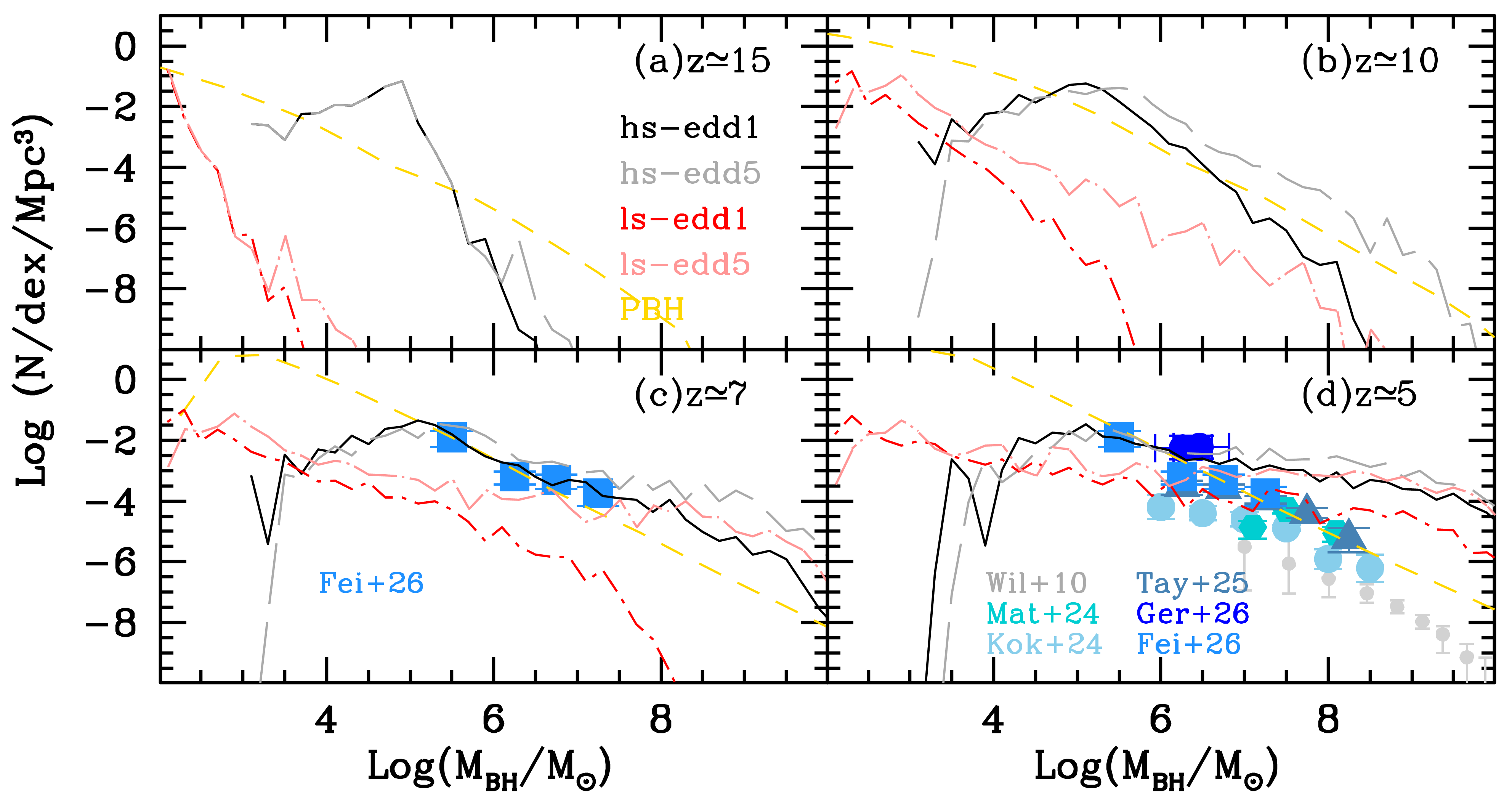}} 
\caption{The redshift evolution of the black hole mass function at $z \sim 5-15$, as marked in panels (a)-(d). Points show observational data from: Wil+10 \citep{willott2010}, Mat+24 \citep{matthee2024}, Kok+24 \citep{kokorev2024}, Tay+25 \citep{taylor2025a}, Ger+26 \citep{geris2026} and Fei+26 \citep{fei2026}. In each panel, lines show theoretical results for the models noted in panel (a) and detailed in Table \ref{table1}.}
\label{fig_bhmf}
\end{center}
\end{figure*}

\subsection{PHANES analytic model for primordial black holes}
\label{phanes}
PBHs generated soon after the Big Bang 
\citep{hawking1971, carr1974, carr2005, carr-green2024} can act as the ``seeds of structure formation". In this scenario, the Coulomb effect of a single black hole generates an initial density fluctuation that can grow through gravitational instability \citep[e.g.][]{hoyle1966, carr-rees1984, carr-silk2018}. Such PBHs have already shown to naturally yield high $M_{\rm BH}-M_*$ ratios \citep{dayal2024_PBH,zhang2025, ziparo2025} and low-metallicity values for the host \citep{dayal2026_PBH} that are generally incompatible with astrophysically-motivated black hole seeding mechanisms \citep{maiolino2025}. 

We use a power-law form of the PBH mass spectrum between $10^{0.5-6}\msun$ such that ${dn}/{dm} = \kappa m^{-\alpha}$ \citep{carr1975, harada2016}, where we use a value of $\alpha = 2$. The normalisation, $\kappa$, is determined using two of the highest redshift black hole candidates, UHZ1 and GHZ9 at $z \sim 10$ and $10.4$, respectively, that are inferred to have a number density of $10^{-5.27} {\rm cMpc^{-3}}$ \citep{kovacs2024, bogdan2024} and a PBH seed mass of $10^{3-4-3.9} \msun$ \citep{dayal2024_PBH}. In this model, individual PBHs start by assembling dark matter linearly around themselves \citep{carr-silk2018}. Assuming this process starts at $z_{\rm mreq}=3400$, the redshift of matter-radiation equality, by redshift $z_{\rm h}=34$, a PBH of mass $M_{\rm PBH}$ can bind a halo mass $M_h$ which is two orders of magnitude larger than the PBH mass. At this point, the halo comes to dominate the gravitational potential - we assume the halo 
then starts growing non-linearly by accreting dark matter from the intergalactic medium (IGM). To estimate these growth rates, we use the average halo accretion rates from high-resolution N-body (dark matter only) simulations \citep{trac2015}. 

In order to model baryons, we assume only halos with a baryonic over-density of $\delta_b = 200$ (with a corresponding halo mass of $M_h^{\rm minb}$) can bind baryons at any redshift. Once a halo exceeds this value, it attains an initial gas mass of $M_g^i = (\Omega_b/\Omega_m) M_h$, a part of which can be accreted onto the black hole and fuel star formation. The mass accreted by a black hole in a given time-step is calculated as
\begin{eqnarray}
    \Delta M_{BH} \propto 
    \begin{cases}
    \fed \med; ~ {\rm if} f_{\rm Edd}M_{\rm Edd}<M_g^i \\
    \frac{\Delta t}{t_{\rm ff}} M_g^i; ~ {\rm if} \fed \med>M_g^i \\
    \end{cases}
    \label{eq_evolvingIMF}
\end{eqnarray}
Here, $t_{\rm ff}$ is the free-fall timescale for the gas in the halo and we use a constant time-step of $\Delta t=30$ Myrs throughout this work. For the sake of simplicity, we limit our calculations to a black hole spin value of $s=0$; we refer interested readers to \citet{dayal2026_PBH} for results on black holes with spin values of $s=+1$ and $s=-1$. In terms of star formation, at any given time-step, the ``effective" instantaneous star formation efficiency is calculated as $f_*^{\rm eff} = min[(\Delta t/t_{\rm ff}), \fej]$. I.e. the effective star formation efficiency is the minimum between the star formation efficiency that produces enough SNII energy to ``unbind" the remainder of the gas ($\fej$), and the ratio between the time-step and the free-fall time.  We account for the impact of both SNII and black hole feedback on the gas mass. We require weak feedback coupling of both black hole ($f_{\rm BH}^w \sim 10^{-3}$) and star formation energies ($f_*^w<0.01$) to the gas content in order to match to the observed black hole and stellar mass combinations for UHZ1 and GHZ9 \citep{dayal2024_PBH}. Further, we require $f_{\rm Edd}=0.25$ to match to the black hole mass-stellar mass observations of all available systems at $z \sim 5-10.4$; we allow a scatter of 0.5 dex on all three of these quantities.

\section{Redshift evolution of global demographics of early black holes}

In this section we start by discussing the redshift evolution of two of the key global demographics for early black holes, for all the models considered in this work: the black hole mass function and the bolometric luminosity function. 

\subsection{Redshift evolution of the black hole mass function}
\label{sec_bhmf}
We start by discussing the redshift evolution of the black hole mass function (BHMF), shown in Fig. \ref{fig_bhmf}, for all the models considered in this work. Globally, the BHMF shows an increase in both its amplitude and mass range with decreasing redshift as continually larger black holes assemble through mergers and accretion with cosmic time in the astrophysical scenarios, and accretion only in the PBH model. 

In the {\sc delphi} model, black holes are seeded starting at $z \sim 40$ (when the first halos form) down to $z \sim 13$. The number densities of newly formed black hole seeds increase with decreasing redshift due to the continual formation of light halos. By $z \sim 15$, even the earliest black holes have only had about 200 Myr to grow. As a result, by this redshift all of the astrophysical seeding models show the early stages of black hole assembly. Both of the heavy seeding models ({\sc hs-edd1} and {\sc hs-edd5}) show very similar BHMFs that range between $\mbh \sim 10^{3-6}\msun$; the {\sc hs-edd5} has a slight heavy tail extending out to $10^{6.5} \msun$ for the rarest objects with a number density of about $10^{-8} {\rm cMpc^{-3}}$. These BHMFs peak at a mass of about $10^5 \msun$, falling off on either side. With their much lower initial black hole masses, light seeding models ({\sc ls-edd1} and {\sc ls-edd5}) show a peak at about $100 \msun$ and attain maximum masses $\sim 10^{3.5-3.75} \msun$ at a number density of $10^{-8} {\rm cMpc^{-3}}$. With their much longer histories and larger range of seed masses, even at $z \sim 15$, PBHs extend over 6 orders of magnitude in mass ($10^{2-8.5}\msun$), exceeding the range shown by both the heavy and light seeding models.  Their number densities exceed all of the astrophysical seeding models for $\mbh \gsim 10^6 \msun$ at these early epochs - they attain masses of about $10^{7.5}\msun$ for number densities of about $10^{-8} {\rm cMpc^{-3}}$.

By $z \sim 10$, there is a clear difference in black hole growth in the Eddington-limited versus super-Eddington accretion models in both the light and heavy seeding scenarios, specially at the massive end. This is driven by the fact that black holes in halos above the critical halo limit ($\mcritb$) can accrete at the minimum between $10\%$ of the available gas mass, up to the allowed Eddington fraction. Most massive black holes accrete at $\fed>1$ in the super-Eddington models resulting in black holes as massive as $10^8 ~ (10^{9.5})\msun$ for the {\sc ls-edd5} ({\sc hs-edd5}) models at a number density of $10^{-8} {\rm cMpc^{-3}}$. At the massive end, in fact, heavy seeds accreting at the Eddington limit ({\sc hs-edd1} model) yield a BHMF comparable to light seeds that can grow at super-Eddington rates ({\sc ls-edd5}) - for a black hole mass of $10^7\msun$, both these models predict two orders of magnitude fewer black holes than the maximal {\sc hs-edd5} model. With light seeds limited by Eddington accretion, the {\sc ls-edd1} model yields the lower limit to the BHMF, reaching maximum masses of only about $10^{5.5}\msun$. PBHs predict a BHMF that lies between the Eddington-limited and super-Eddington models for heavy seeds for black hole masses larger than $10^7\msun$. However, this model predicts two orders of magnitude larger number densities for intermediate-mass black holes ($10^4 \msun$) compared to any astrophysical seeding model. We remind the reader that the number densities of PBHs have been determined by matching to the masses ($10^{7.5-8}\msun$) and number densities ($\sim 10^{-5.5}~ {\rm cMpc^{-3}}$) of the most massive black holes observed at $z \sim 10$ \citep{bogdan2024, kovacs2024}.

By $z \sim 7$, growing black holes in the heavy-seed astrophysical models ({\sc hs-edd1} and {\sc hs-edd5}) generally accrete a fraction of the available gas mass (since this is lower than the fractional Eddington mass). This results in a convergence of the BHMFs - for example, for a black hole mass of $10^7\msun$, the number densities predicted by these two models are only different by a factor 2. The black holes in these models grow to masses of about $10^{10}\msun$ for number densities of about $10^{-8} {\rm cMpc^{-3}}$. While black holes with $\mbh \lsim 10^7 \msun$ still grow slower in the {\sc ls-edd5} model, for higher masses, that BHMF lies between the heavy Eddington-limited and super-Eddington models and reaches out to $10^{10}\msun$. Again, the {\sc ls-edd1} model shows the lower limit to the growth of light black holes, only reaching maximal masses of about $10^{7.5}\msun$ for a number density of $10^{-8} {\rm cMpc^{-3}}$. Interestingly, the PBH BHMF shows a clear excess - by about two orders of magnitude - of intermediate-mass black holes at $\mbh \sim 10^{3-4}\msun$ compared to any astrophysical seeding model; at $\mbh \sim 10^{5-8}\msun$, it predicts results that are very similar to the heavy seeding models. We also find both the heavy seeding ({\sc hs-edd1} and {\sc hs-edd5}) and PBH models to be in excellent agreement with the observationally-inferred BHMF at $z \sim 7$ \citep{fei2026}. Light seeding models predict a much flatter slope than the observations although the {\sc ls-edd5} model is in accord with data for black holes with $\mbh \gsim 10^{6.5}\msun$; given the slower growth of light-seeded Eddington limited black holes, these observations can already be used to rule out the {\sc ls-edd1} model. We briefly note that given our ``heavy seeds" have a range of masses between $10^{3-5}\msun$, we do not find any specific peak in the BHMF; this is in contrast to the results of \citet{jeon2025_bhmf} who find a well-defined peak at $10^5 \msun$, possibly since all of their heavy seeds are initiated with this fixed mass.  

The fact that an increasing fraction of black holes grow by accreting a fraction of the available gas mass yields a greater convergence between all of the astrophysical seeding models considered here by $z \sim 5$ as shown in panel (d) of the same figure. For example for a black hole mass of $10^7\msun$, the heavy-seed Eddington-limited and super-Eddington models and light-seed super-Eddington models show convergent results; the {\sc ls-edd1} model yields a number density that is lower by only a factor 8.  The BHMFs from these different models all extend between $10^{3.5-10}\msun$ - the most crucial difference lies at the low-mass end where the light seeds models predict a tail of black holes extending down to $10^2\msun$ in increasing numbers. These are black holes starved of growth due to a lack of gas in the low-mass halos they reside in. The PBH model persists in predicting an over-abundance of intermediate-mass black holes (with $\mbh \lsim 10^4 \msun$), by about 3 orders of magnitude, compared to any of the astrophysical seeding models. Whilst the heavy (light) seeding models are in reasonable agreement with the data for black holes with $10^{4.5-5.5} ~ (10^{6-7})\msun$, they all predict the existence of massive black holes ($\mbh \gsim 10^{7.5-8}\msun$) that have yet to be observed. The PBH model yields the closet-match to the bulk of the black holes observed with the JWST, and the slope for the most massive black holes ($\gsim 10^8\msun$) observed by combining data from the Sloan Digital Sky Survey and the Canada-France High-z Quasar Survey \citep{willott2010}.

Globally, as of now, the only model that can be ruled out is the is the {\sc ls-edd1} model (light seeds accreting at the Eddington-limit) since it under-predicts the data at $z \sim 7$. We note observations at $z \sim 4.5-7$ are probing BHMF down to $10^{5.5}\msun$ and are expected, with further lensing observations, to go down to even lower masses. These would provide a definite means of, at least, distinguishing between astrophysical and cosmological seeding models given the over-abundance of the latter at low masses ($\lsim 10^4 \msun$) at $z \sim 5-10$.  

\subsection{Redshift evolution of the bolometric luminosity function (BLF)}
\label{sec_blf}
\begin{figure*}
\begin{center}
\center{\includegraphics[scale=0.12]{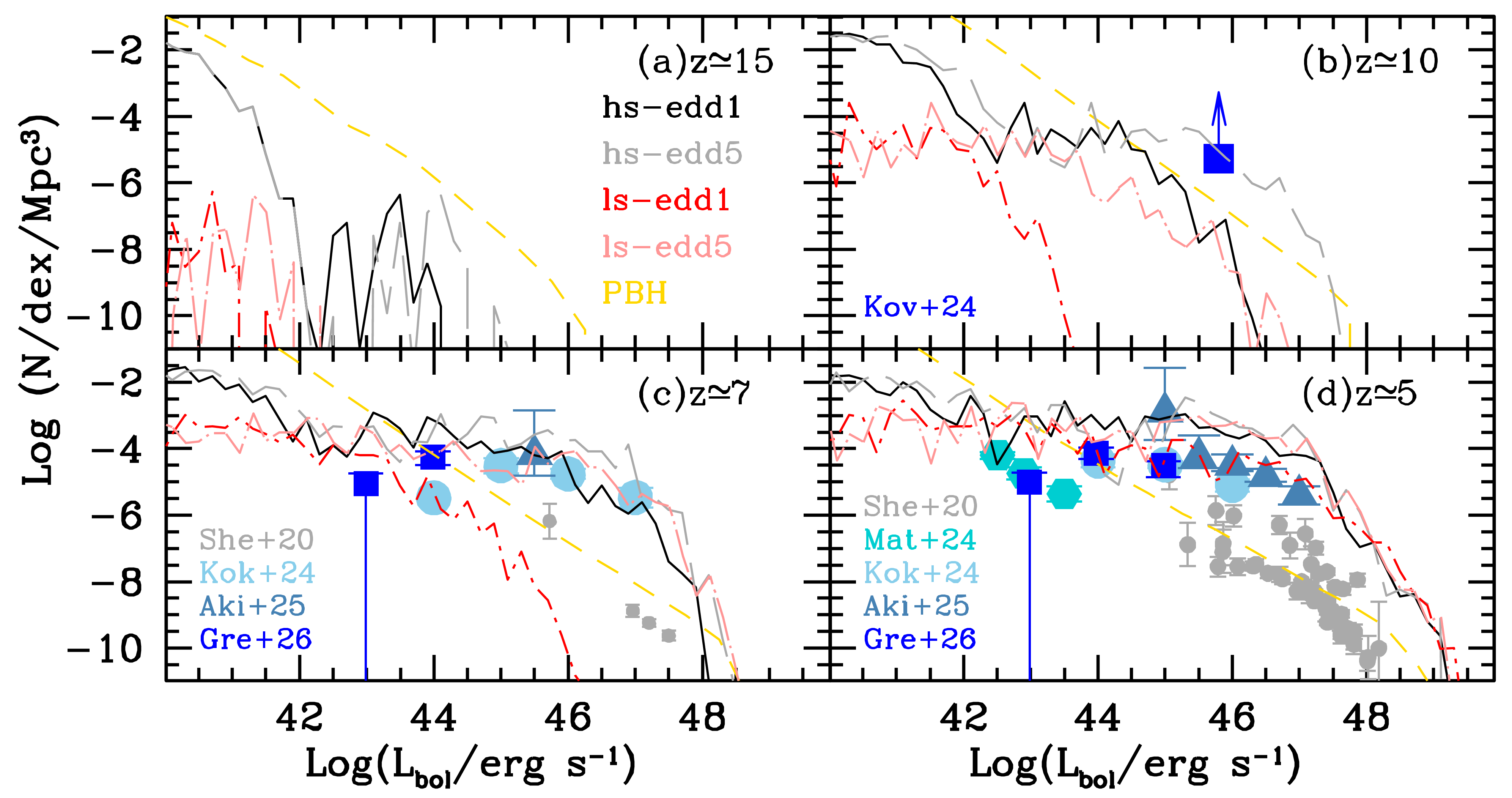}} 
\caption{The redshift evolution of the bolometric luminosity function for black holes at $z \sim 5-15$, as marked. In each panel, the lines show results for the theoretical models marked in panel (a). Points show observational data: at $z \sim 10$, the point shows the lower limit to the bolometric luminosity function based on UHZ1 and GHZ9 from Kov+24 \citep{kovacs2024}. At $z \sim 7$ the data is from from She+20 \citep{shen2020}, Kok+24 \citep{kokorev2024}, Aki+25 \citep{akins2025a} and Gre+26 \citep{greene2026}; at $z \sim 5$ from She+20 \citep{shen2020}, Mat+24 \citep{matthee2024}, Kok+24 \citep{kokorev2024}, Aki+25 \citep{akins2025a} and Gre+26 \citep{greene2026}. }
\label{fig_bollf}
\end{center}
\end{figure*}

We now discuss the redshift evolution of the bolometric luminosity function, the results of which are shown in Fig. \ref{fig_bollf}. In our work, in order to compare with observations, the bolometric luminosity of a black hole (with $\fed>0.01$) is calculated as $\lbol = \epsilon_r\,\Delta \mbh, c^2 \Delta t^{-1}$, where $c$ is the speed of light and $\Delta \mbh$ is the increase in the black hole mass in the time-step of $\Delta t = {\rm 30~ Myr}$. For values of $\fed<0.01$, the bulk of the thermal energy is carried into the black hole rather than being radiated \citep{narayan1994,yuan2014}. As a result, we consider such inefficiently accreting black holes to produce no luminosity - this has no bearing on our results though \citep[for a discussion see][]{caceres2026}. With these calculations, at $z \sim 15$, the BLF from putative heavy (light) seeding models only extends to $\lbol \sim 10^{44-45} ~ (\lsim 10^{42})\ergs$ for a number density corresponding to $10^{-8} {\rm cMpc^{-3}}$. Given their overall larger masses and accretion rates, the PBH BLF dominates at all $\lbol \sim 10^{40-46.5} \ergs$ at this redshift and shows bolomteric luminosities as high as $10^{45.5}\ergs$ for a number density of $10^{-8} {\rm cMpc^{-3}}$. By $z \sim 10$, the growth of black hole seeds leads to a corresponding increase in the BLF: while heavy seeding models converge for $\lbol \lsim 10^{44} \ergs$, the fast growth of massive seeds in super-Eddington model ({\sc hs-edd5}) results in it yielding the upper limit to the BLF. For a number density of $10^{-8} {\rm cMpc^{-3}}$, this model predicts $\lbol \sim 10^{47.5}\ergs$ compared to the 30-times lower value of $10^{46}\ergs$ in the {\sc hs-edd1} model. Accreting at super-Eddington rates, black holes in the {\sc ls-edd5} model show a very similar BLF compared to the {\sc hs-edd1} model, specially at the bright-end. With Eddington-limited growth onto light seeds, the {\sc ls-edd1} model yields the lower limit to the BLF with maximum values as low as $\lbol \sim 10^{43}\ergs$ for $10^{-8} {\rm cMpc^{-3}}$. With its over-abundance of intermediate mass black holes, the PBH BLF exceeds the amplitude of any of the astrophysical seeding models: for $\lbol \sim 10^{43} \ergs$, the PBH model predicts two orders of magnitude more black holes compared to any of the astrophysical seeding models. We also compare to the lower-limit on the BLF from two sources (UHZ1 and GHZ9) inferred at this redshift \citep{kovacs2024}. As seen, the heavy seeding, super-Eddington accretion model ({\sc hs-edd5}) is the only one in accord with this data point. Although in agreement with the mass estimates, the PBH model lies slightly below this point. This is because observations assume Eddington-limited accretion to convert between the black hole mass and its bolometric luminosity - however, PBHs accrete at sub-Eddington limits ($\fed \lsim 0.25$) in our model. Confirmation of such high number densities of luminous black holes at these early epochs would allow ruling out all light-seeding models, in addition to Eddington-limited heavy-seeding ones.

By $z \sim 7$, mimicking the convergence of the BMHFs, both the heavy seeding models and the light-seed super-Eddington models start showing very similar BLFs for $\lbol \gsim 10^{42}\ergs$. We note that the parameters of the {\sc hs-edd1} model have been tuned to reproduce observations and we have used the same parameters for all of the astrophysical seeding models studied here. This convergence in theoretical models results in both the heavy seeding models ({\sc hs-edd1} and {\sc hs-edd5}) and the light super-Eddington model ({\sc ls-edd5}) being in excellent accord with the JWST-observed BLF at this redshift, at least for $\lbol \gsim 10^{44}\ergs$. The {\sc ls-edd1} model again yields the lower-limit to the BLF and under-predicts the bright end of the observed BLF given the slower black hole growth (and associated low luminosities) in this model. Finally, we note that the PBH BLF again over-predicts the low-luminosity end ($\lbol \lsim 10^{42}\ergs$) by about two orders of magnitude compared to all the other models studied here; it somewhat under-predicts the bright end of the BLF inferred from the JWST at this redshift, although it is in excellent accord with ground-based results \citep{willott2010}. 

\begin{figure*}
\begin{center}
\center{\includegraphics[scale=0.12]{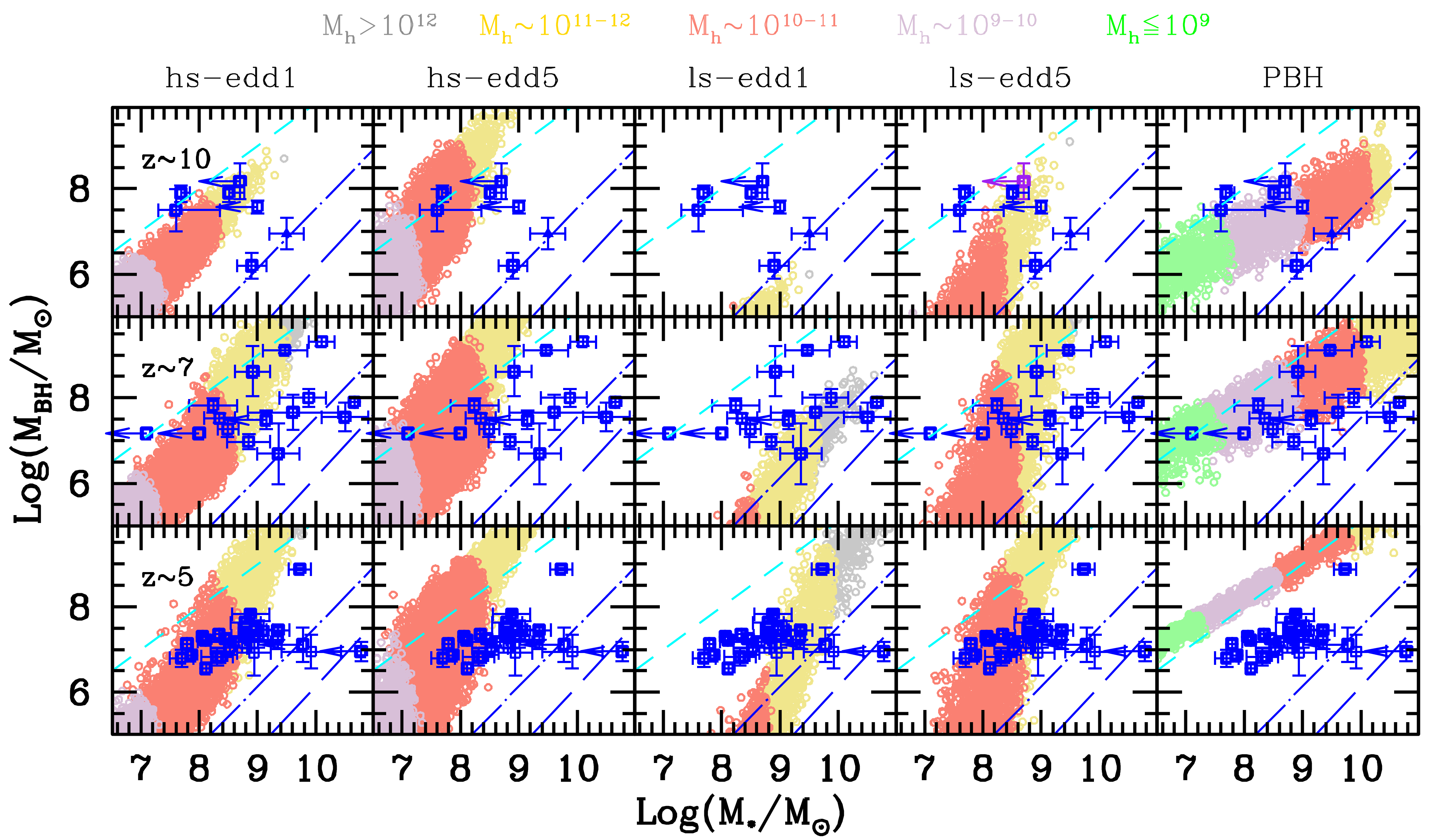}} 
\caption{The redshift evolution of the black hole mass-stellar mass relation. Rows show results at $z \sim 5-10$, with each column referring to the theoretical model marked above. In each panel, the colours refer to the host halo mass associated with the object, as marked above the figure. At each redshift, we show observational results from a number of groups: at $z \sim 5$ from \citet{harikane2023bh}, \citet{maiolino2024_jades}, \citet{kocevski2023, kocevski2025}, \citet{matthee2025}; at $z \sim 7$ from \citet{harikane2023bh}, \citet{maiolino2024_jades}, \citet{furtak2024}, \citet{tripodi2025}, \citet{juodzbalis2024}, \citet{kocevski2025}, \citet{akins2025}, \citet{maiolino2025}; and $z \sim 10$ from \citet{kokorev2023}, \citet{larson2023}, \citet{bogdan2024}, \citet{kovacs2024}, \citet{maiolino2025}, \citet{Napolitano2025_Xray}, \citet{taylor2025}. In all panels, dot-dashed and long-dashed blue lines show observationally-inferred local relations from \citet{reines2015} and \citet{suh2020}, respectively; in each panel the short-dashed cyan line shows a relation where the black hole mass equals the stellar mass.}
\label{fig_mbhfnz}
\end{center}
\end{figure*}

By $z \sim 5$, within errors, the BLFs for all of the astrophysical models converge for $\lbol \gsim 10^{42} \ergs$ given the convergence of both the BHMFs and the fact that most black holes grow by accreting a certain fraction of the available gas mass. These models all predict a value of $\lbol \sim 10^{48} \ergs$ for a number density of $10^{-8} {\rm cMpc^{-3}}$. Given the scatter of the observed BLF, these models all reasonably reproduce the observations at $\lbol \gsim 10^{44}\ergs$. Although still over-predicting the low-luminosity end of the BLF ($\lbol \lsim 10^{42}\ergs$), the low Eddington rates ($\fed \lsim 0.25$) of PBHs results in this model yielding the lower limit to the BLF for luminous systems. However, we note we have used a continuous slope of $\alpha =-2$ for the PBH mass function - a shallower slope would alleviate this issue.

Overall, only the {\sc hs-edd5} and PBH models are in accord with the data at $z \sim 10$. At $z \sim 5-7$ all astrophysical models, apart from {\sc ls-edd1}, are in reasonable agreement with observations within the bounds of the scatter and errors; the amplitude of the {\sc ls-edd1} model is too low compared to observations at $z \sim 7$, ruling it out. Finally we caution that recent results from \citep{greene2026} find that more than half of the bolometric luminosity for two little red dots (at $z=3.1$ and 4.46) emerges in the rest-frame optical with the X-ray corona, UV-emitting blackbody and mid-far infrared emissions being considerably subdominant. The new bolometric corrections in this work lower the inferred luminosities by a factor of 10 compared to published values in the literature. Complete SEDs of a larger sample of early black holes are crucially needed to ascertain the true bolometric luminosities of early black holes and differentiating between theoretical models.  

\section{Physical properties of early black holes and their hosts }
We now discuss the detailed physical relations between black holes and their hosts, in the different models studied in this work, in what follows.

\subsection{Relation between black holes and host stellar masses}
\label{sec_mbhms}
We start by discussing the black hole-stellar mass relations at $z \sim 5-10$, color-coded by the host halo mass. We reiterate that both the black hole and stellar masses of these high-redshift sources remain extremely poorly constrained, as discussed in the introduction. With these caveats in mind, we compare theoretical results to observations where we take the latter at face value for the time being. We start by noting a positive correlation between the black hole, stellar and halo masses, irrespective of the model considered. Further, while the black hole, stellar and halo masses grow with time both through mergers and accretion in the astrophysical seeding models, they grow through accretion alone in the PBH model. The black hole mass-stellar mass relation for each model is tabulated in Table \ref{table2}.

Starting at $z \sim 10$, we find black holes as massive as $\mbh \sim 10^{8-9}\msun$ already in place in the {\sc hs-edd1} model. Such massive black holes inhabit galaxies with $M_* \sim 10^{8-9.5}\msun$ and halos of mass $\mh \sim 10^{11-12}\msun$. The $\mbh-M_*$ relation in this model starts approaching the 1-to-1 relation given the growth of heavy seeds in putative stellar systems. This model can explain the bulk of the systems observed, apart from the most extreme outliers that imply $\mbh \sim M_*$ \citep{kovacs2024}. As expected, the faster growth of heavy seeds in the {\sc hs-edd5} model results in a steeper $\mbh-M_*$ slope, with over-massive black holes that exceed the stellar mass for all $\mbh \gsim 10^6\msun$. This model yields the upper limit to the $\mbh-M_*$ relation in astrophysical seeding models and shows black holes as massive as $10^{9-10}\msun$ in galaxies with $M_* \sim 10^{8-9}\msun$ and halos of mass $\mh \sim 10^{11-12}\msun$. The {\sc ls-edd1} model naturally yields the lower-limit to this relation, with the most massive black holes only reaching masses of $10^6 \msun$ in halos as massive as $10^{11-12}\msun$ with $M_* \sim 10^{8-9}\msun$. The results of this model lie between the two local relations shown \citep{reines2015, suh2020} and are completely unable to produce black holes more massive than $0.1\% M_*$. With fast, super-Eddington growth of light seeds, the {\sc ls-edd5} model has the steepest slope of the astrophysical seeding models. In this model, by $z \sim 10$, however, black holes with $\mbh \sim 10^{5-7} \msun$ are hosted in galaxies with $M_* \sim 10^{7-8.5}\msun$ and $\mh \sim 10^{10-11}\msun$ i.e. ratios of $\mbh/M_*\lsim 0.1$. These significantly under-predict the $\mbh \sim M_*$ observational values obtained by e.g. \citet{bogdan2024} and \citet{kovacs2024}. In this model, the most massive systems reach masses of $10^{8.5}\msun$ and show values of $\mbh \sim M_*$ in accord with the high values shown by observations \citep[from][]{kokorev2023,taylor2025}. Finally, the PBH model shows the shallowest slope given that black holes seeding and growth precedes the baryonic component. In this case, black holes of $\mbh \sim 10^{8-10}\msun$ are hosted in galaxies with $M_* \sim 10^{9-10.5}\msun$ and $\mh \sim 10^{10-12}\msun$. As detailed in our previous works \citep{dayal2024_PBH, dayal2026_PBH}, PBH models naturally predict very high $\mbh-M_*$ ratios resulting in excellent agreement with the observations that yield ratios of $\mbh/M_* \gsim 0.3$, even though the black holes in this model can accrete at $\fed \sim 0.25$ at most. 

\begin{table}
\centering
    \begin{tabular}{|c c c c|}
    \hline
     z & Model & $\gamma$ & $\delta$\\
    \hline
  5 & {\sc hs-edd1} & $1.19$ & $-8.66$ \\
   5 & {\sc hs-edd5} & $2.18$ & $-9.75$ \\
   5 & {\sc ls-edd1} & $2.49$ & $-16.27$ \\
   5 & {\sc ls-edd5} & $2.85$ & $-17.31$ \\
   5 & {\sc PBH} & $0.08$ & $+1.84$ \\
   7 & {\sc hs-edd1} & $1.52$ & $-5.53$ \\
   7 & {\sc hs-edd5} & $1.80$ & $-10.54$ \\
   7 & {\sc ls-edd1} & $2.15$ & $-9.46$ \\
   7 & {\sc ls-edd5} & $2.62$ & $-15.66$ \\
   7 & {\sc PBH} & $0.58$ & $+3.09$ \\
   10 & {\sc hs-edd1} & $1.19$ & $-2.69$ \\
   10 & {\sc hs-edd5} & $0.98$ & $-3.55$\\
   10 & {\sc ls-edd1} & $1.83$ & $-6.73$\\
   10 & {\sc ls-edd5} & $1.86$ & $-9.76$\\
   10 & {\sc PBH} & $0.61$ & $+1.95$\\
          \hline
    \end{tabular}
    \caption{For the redshift (column 1) and model (column 2), we quote the relation between the black hole mass and stellar mass quantified as ${\rm log}(\mbh) = \gamma {\rm log}(M_*) +\delta$. We limit this to a minimum stellar mass of $10^7\msun$.}
    \label{table2}
\end{table}

By $z \sim 7$, black holes with $\mbh \gsim 10^{6.5}\msun$ are able to grow larger than the stellar mass in the {\sc hs-edd1} model. Such black holes inhabit galaxies with $M_* \lsim 10^7 \msun$ and $\mh \sim 10^{9-10}\msun$, with lower mass black holes showing ratios of $\mbh/M_* \sim 0.3$. This model is in accord with all of the observations, including systems with $\mbh/M_* \sim 1$; the only data it is unable to explain concerns black holes with $\mbh \sim 10^7 \msun$ that are over-massive with respect to their hosts, with $M_* <10^7\msun$ \citep{maiolino2024_jades}. A fraction of black holes of all masses, $\mbh \sim 10^{5-10}\msun$ show values of $\mbh/M_*>1$ in the {\sc hs-edd5} model. Interestingly, massive black holes, with $\mbh \gsim 10^{8-9.5}\msun$ are hosted in halos of $10^{10-11}\msun$ in this model while they are hosted in much larger halos ($10^{11-12}\msun$) for the {\sc hs-edd1} model given their slower growth. This model is not only in accord with all the observations but predicts a population of extremely obese black holes at all masses. This implies only a small fraction of all halos can contain heavy seeds that can accrete at super-Eddington rates to be in accord with current high-$z$ observations. The {\sc ls-edd1} model again yields the lower limit to the black hole growth with $\mbh/M_*<1\%$ for any of the masses here with a maximum black hole mass of $10^{8.5}\msun$. The results of this model are in good accord with the low-redshift relation inferred by \citet{suh2020} and are unable to explain the bulk of observations at $z \sim 7$. These black holes are also typically hosted in halos that are much more massive compared to the other models - for example, $\mbh \sim 10^{8} \msun$ black holes are hosted in halos larger then $10^{12}\msun$. Continuing to yield the steepest $\mbh-M_*$ relation, we find a fraction of the black holes with $\mbh \gsim 10^{8}\msun$ to be more over-massive than their stellar component in the {\sc ls-edd5} model; they occupy massive halos with $\mh \sim 10^{10-12}\msun$. Although explaining the bulk of the observations, this model too fails to explain the light obese black holes with $\mbh \sim 10^7 \msun$ and $M_* <10^7\msun$ \citep{maiolino2024_jades}. Finally, the PBH model again shows the shallowest $\mbh-M_*$ slope, predicting $\mbh/M_* \sim 0.3-1.5$ at all mass ranges, as in the {\sc hs-edd5} model; this model naturally explains all of the observations available so far. A key difference between these models, however, is that given the co-evolution of the black hole and its halo in PBH models, black holes of a given mass typically occupy halos that are less massive compared to any of the astrophysical seeding models. For example, black holes of $10^{6-8}~ (10^{8-9})\msun$ occupy halos with $\mh \lsim 10^{9} ~ (10^{9-10})\msun$, a point we discuss in more detail in Sec. \ref{sec_hosts}. 

\begin{figure*}
\begin{center}
\center{\includegraphics[scale=0.12]{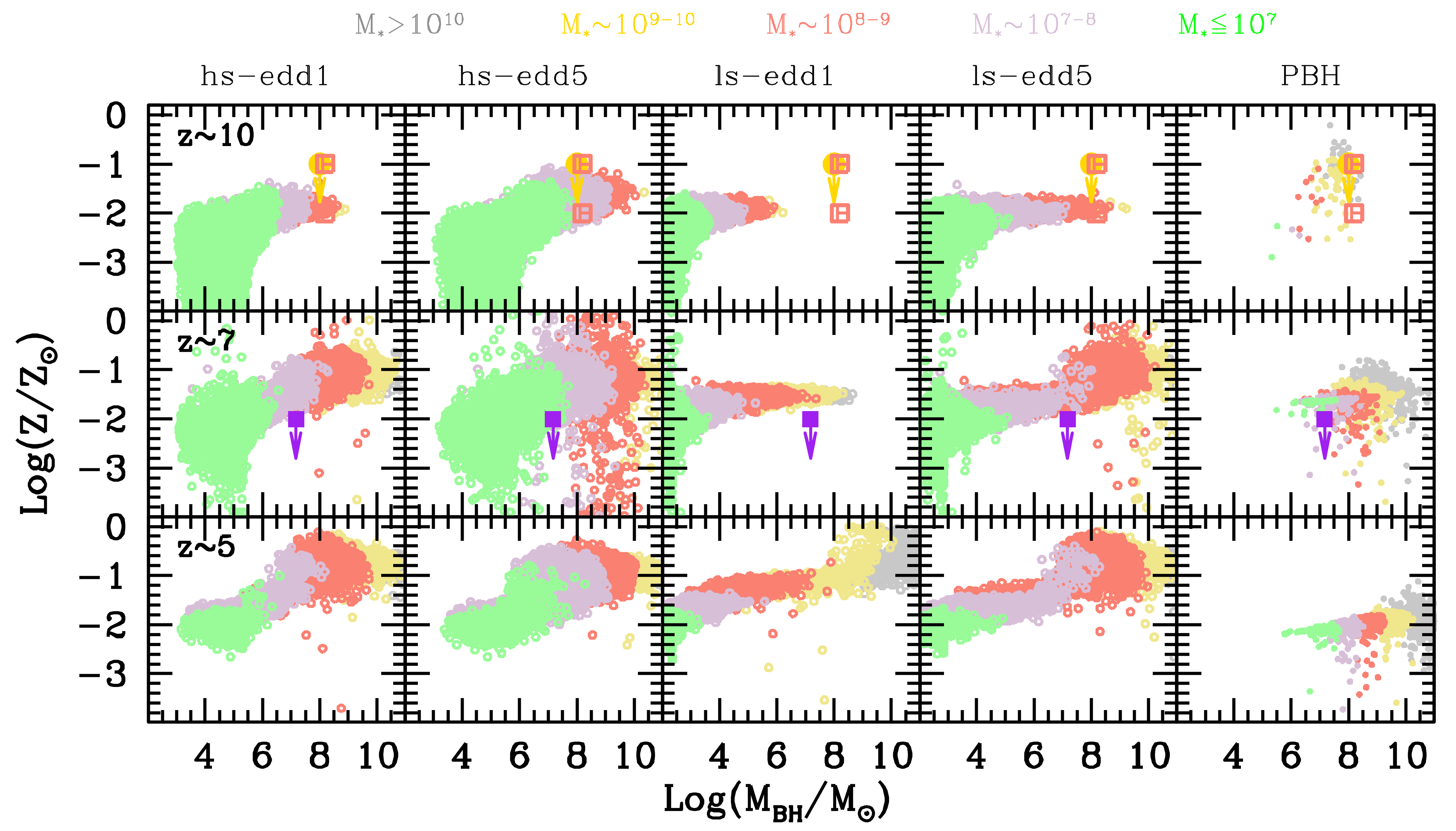}} 
\caption{The redshift evolution of the gas-phase metallicity as a function of the black hole mass. The rows shows results at $z \sim 5-10$, with each column referring to the theoretical model marked above. In each panel, the colours refer to the stellar mass associated with the object, as marked above the figure. The filled points at $z \sim 7$ and $z \sim 10$ show upper limits on the metallicity estimate for Abell2744-QSO1 from \citet{maiolino2025} and CANUCS-LRD-z8.6 at $z \sim 8.6$ from \citet{tripodi2025}, respectively. Finally, the empty data points at $z \sim 10$ shows the metallicities estimated for GHZ9 by \citet{Napolitano2025_Xray}.}
\label{fig_metbh}
\end{center}
\end{figure*}

The evolution down to $z \sim 5$ yields very similar results as at $z \sim 7$: the heavy seeding models start showing very similar results since black holes generally accrete a certain fraction of the available gas mass in both cases. The key difference arises for $\mbh \lsim 10^8\msun$ black holes which are still less massive than the stellar mass in the {\sc hs-edd1} model while black holes of all masses show $\mbh/M_*>1$ for the {\sc hs-edd5} model. At this redshift, $\mbh \sim 10^8 \msun$ black holes are hosted in galaxies with $M_* \sim 10^{7.8-9}\msun$ and $\mh \sim 10^{10-11}\msun$ in both models; both these models are in accord with all of the observations at this $z$. While black holes are able to grow up to $10^{10}\msun$ in the {\sc ls-edd1} model, they are always less massive than the stellar mass - they are again hosted by larger halos than any of the other astrophysical seeding models and crucially, predict much lower black hole masses than observed for a bulk of the observations, specially those with $\mbh \sim 10^7 \msun$ and $M_* \lsim 10^9 \msun$ \citep{harikane2023bh,maiolino2024_jades,kocevski2025}. The {\sc ls-edd5} model yields very similar results to the heavy seeding models, specially at $\mbh \gsim 10^8\msun$, in terms of the underlying stellar and host halo masses. Finally, most of the PBH-seeded black holes lie on or above the $\mbh = M_*$ relation. We again note that black holes of a given mass are hosted in extremely light halos in this model. For example, $\mbh \sim 10^8 \msun$ objects are hosted in galaxies with $M_* \sim 10^{7.4-8.2}\msun$ with $\mh \sim 10^{9-10}\msun$. This model provides an upper limit to the observations at this redshift.

Overall, we find that the heavy seeding super-Eddington ({\sc hs-edd5}) and the PBH models are the only ones capable of explaining the extremely high black hole to stellar mass ratios at {\it all} redshifts $z \sim 5-10$. While the  heavy seeding Eddington-limited model can reproduce the bulk of the objects, it can not explain the existence of black holes with $\mbh \sim 10^7 \msun$ that are over-massive with respect to their hosts, with $M_* <10^7\msun$, at $z \sim 7-10$.

\subsection{The metallicity evolution of early black holes}
\label{sec_metmbh}
We now discuss the relation between the gas-phase metallicity of the host and the black hole mass, colour-coded as a function of the stellar mass, in Fig. \ref{fig_metbh}. Recent observations have yielded metallicities of 3 black hole hosts between $z \sim 7-10$ \citep{maiolino2025, tripodi2025,Napolitano2025_Xray}. All of these works use different methodologies to infer the metallicity: \citet{maiolino2025} infer an extremely low value of $4\times 10^{-3} \zsun$ based on the weakness of the [OIII]5007 emission line relative to the narrow H$\beta$ emission. \citet{Napolitano2025_Xray} use Ne3O2 ratio-based calibrations to derive a value $\sim 0.01-0.1 \zsun$ and \citet{tripodi2025} use direct electron temperature methods to derive $Z \lsim 0.1 \zsun$. 

Theoretically, the gas-phase metallicity scales with the stellar mass - and hence the black hole mass given the $\mbh-M_*$ relation discussed in Sec. \ref{sec_mbhms} above; the slope of the $Z-\mbh$ relation naturally depends on the exact model considered. At $z \sim 10$, in the {\sc hs-edd1} model, light black holes ($\lsim 10^6\msun$) are hosted in putative galaxies ($M_* \lsim 10^7 \msun$) which have low values of metallicity ($Z \lsim 0.03 \zsun$). This is because the small amount of metals produced by stars are either astrated into further star formation/black hole accretion or lost due to feedback. The metallicity value increases only very slightly with black hole mass - this is because the metals produced by star formation are balanced by those lost through astration and outflows. In this model, the maximum metallicity value is about $0.05~\zsun$ for $\mbh \sim 10^{8-9}\msun$ hosted in galaxies with $M_* \sim 10^{8-10}\msun$. Black holes naturally build mass faster in the {\sc hs-edd5} model - as a result this model shows a larger scatter in the $\mbh-Z$ relation as compared to the {\sc hs-edd1} model. Here too, the metallicity in light black hole systems ($\mbh \lsim 10^6\msun$) is less than a few percent of the solar value and shows a peak at $\sim 0.1 \zsun$ for $\sim 10^7 \msun$ back holes hosted in $M_* \sim 10^{7-8}\msun$ systems. It settles back to about $0.03~\zsun$ for more massive black holes given metal loss in astration and feedback. With the smallest mass range for black hole growth, the {\sc ls-edd1} model shows $Z \lsim 0.02~\zsun$. While light seeds grow up to $10^9\msun$ allowing super-Eddington accretion, their metallicity trend is rather flat for $\mbh \sim 10^{5-9}\msun$ showing $Z \sim 0.01 \zsun$. Finally, given their longest assembly histories and low-accretion and star formation rates, PBHs are as metal enriched as $0.1 \zsun$ even at $z \sim 10$. We also compare these results to the observations from \citet{tripodi2025} and \citet{Napolitano2025_Xray}. 
Only the {\sc hs-edd1} and {\sc ls-edd5} models yield black hole masses, stellar masses and metallicities in agreement with the lower metallicity value ($0.01 \zsun$) inferred by \citet{Napolitano2025_Xray}. However, none of the astrophysical seeding models yield the right combination of black hole mass, stellar mass and metallicity compared to \citet{tripodi2025}: while the {\sc hs-edd1}, {\sc hs-edd5} and {\sc ls-edd5} models show metallicity and black hole masses in agreement with the observations, they predicts a lower stellar mass ($10^{7-9}\msun$) as compared to the observations, where $M_* \sim 10^{9-10}\msun$. The PBH model is the {\it only} one that reproduces the observations along all three axes, at least compared to \citet{tripodi2025}. Even this model predicts a slightly different stellar mass ($10^{9-10}\msun$) compared to the value of $10^{8-9}\msun$ obtained by \citet{Napolitano2025_Xray}; these authors however caution the stellar mass can vary by a factor 2 depending on the black hole contribution to the SED.  

The growth of both the black hole and stellar components by $z \sim 7$ results in a corresponding build-up of the metal mass for all models. The metallicity essentially scales with the black hole mass for the {\sc hs-edd1}, {\sc hs-edd5} and {\sc ls-edd5} models, albeit with varying degrees of scatter. In the heavy seeding models, systems with $\mbh \sim 10^7 \msun$ ($M_* \sim 10^{7-8}\msun$) show $Z \sim 0.1 \zsun$, increasing to $Z \gsim 0.5 \zsun$ for $\mbh \sim 10^9\msun$ ($M_* \sim 10^{8-9}\msun$). Given the fast growth of heavy seeds, the {\sc hs-edd5} model shows the largest scatter in the black hole mass-metallicity relation. The metallicity trend is again quite flat for the {\sc ls-edd1} model which shows $Z \lsim 0.1 \zsun$ for all black hole masses $\mbh \sim 10^{2-8.5}\msun$. The {\sc ls-edd5} model yields similar results to the {\sc hs-edd1} model, with $Z \sim 0.1 ~ (\gsim 0.5)\zsun$ for $\mbh \sim 10^7 ~ (10^8)\msun$. Finally, the PBH model shows metallicity values that increase from about $0.03 - 0.1 \zsun$ as the black hole mass increases from $10^7 - 10^9 \msun$ with associated values of $M_* \sim 10^{9-10}\msun$. At this redshift, we compare to the observations from \citet{maiolino2025} who infer an upper limit of $0.01\zsun$ to the metallicity for a system with $\mbh \sim 10^{7.1} \msun$ and $M_* \sim 10^{7.1-8}\msun$. As seen, the only two models that are in agreement along all three axes are {\sc hs-edd1} and the PBH model. 

By $z \sim 5$, all astrophysical seeding models show similar black hole mass-metallicity relations where the metallicity has a value of $0.1 ~ (\gsim 0.5)\zsun$ for $\mbh \sim 10^7 ~ (10^8) \msun$. Given that PBH halos grow their halos around themselves, black holes of a given mass reside in the smallest potentials in this model. As a result, the metals in this model are more susceptible to feedback and reach maximal values of $Z \sim 0.01 ~ (0.1) \zsun$ for $\mbh \sim 10^7 ~ (10^{10})\msun$ which are hosted in galaxies of $M_* \lsim 10^7 ~ (\sim 10^{9-10})\msun $.

Overall, only the {\sc hs-edd1} and PBH models are in broad accord with the combinations of metallicity, black hole mass and stellar mass available at $z \sim 5-7$, albeit they do {\it not} reproduce all of the observed data points simultaneously.  

\subsection{The host halos of early black holes}
\label{sec_hosts}
\begin{figure*}
\begin{center}
\center{\includegraphics[scale=0.12]{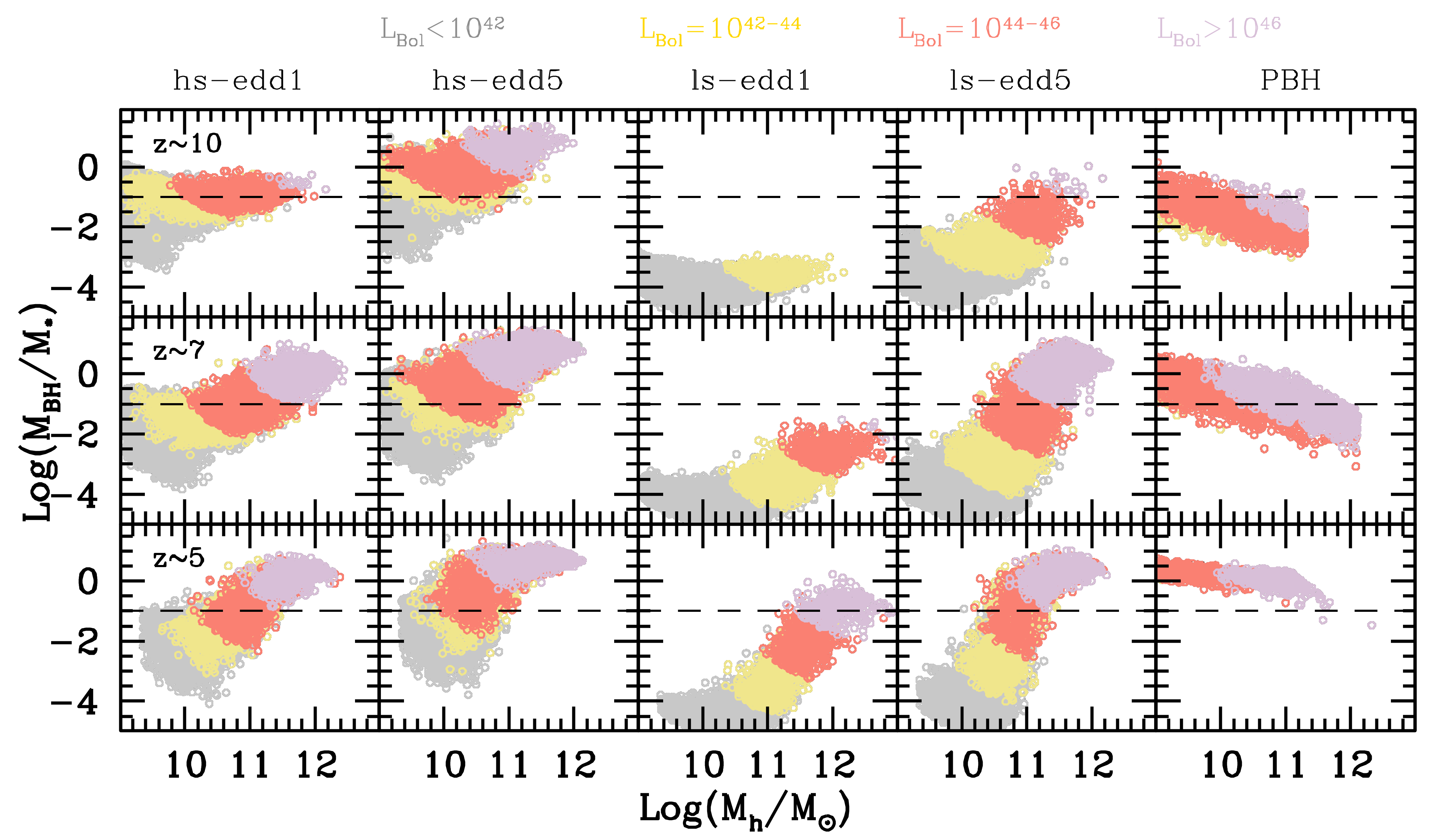}} 
\caption{The redshift evolution of the black hole-to-stellar mass ratio as a function of the halo mass. The rows shows results at $z \sim 5-10$, with each column referring to the theoretical model marked above. In each panel, the colours refer to the black hole bolometric luminosity associated with the object, as marked above the figure. In each panel, the horizontal dashed line shows a value of $\mbh/M_* = 0.1$.}
\label{fig_clus}
\end{center}
\end{figure*}

As seen from Sec. \ref{sec_mbhms} above, one of the strongest differences between astrophysical and cosmological black hole seeding models lies in the halo masses they inhabit. We now compare the black hole-stellar mass ratio as a function of the underlying host halo mass, for different bolometric luminosity limits, as shown in Fig. \ref{fig_clus}. Our key aim is to look for observational imprints that can help distinguish between the different seeding and growth models studied in this work.

We select systems with a ratio of $\mbh/M_* > 0.1$ and $\lbol \sim 10^{44-46} \ergs$ as a {\it gedankenexperiment}. We start by noting that, given the suppressed growth of Eddington-limited light seeds, black holes in the {\sc ls-edd1} model never fulfil these requirements. We also note that in astrophysical seeding models, the black hole to stellar mass ratio increases with increasing host halo mass. This is driven by the deeper potentials and larger gas masses in increasingly massive halos that allow copious amounts of star formation, and accretion onto the black hole. However, the black hole to stellar mass ratio {\it decreases} with increasing halo mass in PBH models where the black hole pre-dates both the halo and stellar mass. Light PBHs are able to assemble light halos that are inefficient at star formation, pushing up the $\mbh/M_*$ ratio. As PBHs grow larger halos around themselves, these can support increasing amounts of star formation - the resulting large stellar masses lead to a decrease in the $\mbh/M_*$ value.

For the heavy seeding models, at $z \sim 10$ systems with $\mbh/M_* > 0.1$ and $\lbol \sim 10^{44-46} \ergs$ inhabit halo masses of $10^{10-12}$ and $10^{9-11.5}\msun$ for the {\sc hs-edd1} and {\sc hs-edd5} models, respectively. Given the small seed masses, there are only a few systems that fulfil these requirements as early as $z \sim 10$ in the {\sc ls-edd5} model - these inhabit $\mh \sim 10^{10.8-11.4}\msun$ halos.  In the PBH model, such systems inhabit halos with $\mh \sim 10^{9-10}\msun$.

By $z \sim 7$, both the heavy seeding and light seeds super-Eddington models start looking very similar: the systems under consideration reside in halos of $10^{10-12}\msun$ for both the {\sc hs-edd1} and {\sc ls-edd5} models; they reside in slightly lower mass halos, down to $10^{9.2-12}\msun$ for the {\sc hs-edd5} model. Finally, in the PBH model such systems inhabit $\mh \sim 10^{9-11}\msun$ halos. 

The results for the three astrophysical models converge by $z \sim 5$ where such systems are hosted in $\mh \sim 10^{10-12}\msun$ halos for the {\sc hs-edd1}, {\sc hs-edd5} and {\sc ls-edd5} models. PBH-seeded systems fulfilling the required criteria reside down to lower-mass halos with $\mh \sim 10^{9-11}\msun$.

Overall, we find that it will be hard to distinguish between different astrophysical seeding and growth models based on clustering given the similarity of halo mass ranges for the chosen selection criterion. However, at $z \sim 7$, selecting systems with $\mbh/M_* > 0.1$ and $\lbol \sim 10^{44-46} \ergs$ that show a {\it negative} black hole to stellar mass ratio and reside in $10^{9-11}\msun$ halos can be used as possible indicators of PBH seeding models. 

JWST observations have been building increasing samples of luminous quasars with $\mbh \sim 10^{9-10}\msun$ with bolomteric luminosities $\gsim 10^{46}\ergs$ at $z \gsim 6$. These have already been used to calculate auto-cross correlation scales of $15-22 h^{-1}{\rm cMpc}$ and associated halo masses of $\mh \gsim 10^{12.2-12.4}\msun$ \citep{eilers2024, wang2026_aspire}. It is therefore extremely conceivable that the experiments we propose - of observing a statistical sample of lower mass black holes, with $\lbol \sim 10^{44-46}\ergs$ at $z \sim 7$ - can be carried out within the next few years. Such observations will be crucial in shedding light on the seeding and growth scenarios of these early systems.
\section{Conclusions and discussion}
\label{sec_conc}
In this work, we have compared the observational properties of astrophysically- and cosmologically-seeded black holes in the first billion years of the Universe. We use the {\sc delphi} semi-analytic model for light ($\sim 100 \msun$) and heavy ($\sim 10^{3-5}\msun$) astrophysical seeds, and the {\sc phanes} analytic model for primordial black holes (PBHs; $\sim 10^{0.5-6}\msun$). Our key findings are as follows:

\begin{itemize}
\item All models predict a black hole mass function that grows both in amplitude and mass range with decreasing redshift as larger black holes assemble through accretion and mergers in the astrophysical models, and accretion alone in the PBH scenario. PBHs predict an overabundance of intermediate-mass black holes ($\sim 10^{3-4}\msun$) by 2-3 orders of magnitude relative to astrophysical models at $\sim 5-10$, whilst remaining consistent with observations at the massive end. The {\sc ls-edd1} model (light Eddington-limited seeds) is the only astrophysical model that can be ruled out given its under-prediction of the observed BHMF at $z \sim 7$. We caution the PBH mass spectrum has been calibrated using two black holes at $z \sim 10-10.4$. Estimates of the PBH mass spectrum from first principles are crucially required in order to test these cosmological scenarios.

\item The bolometric luminosity function too builds up with time as increasingly larger black holes assemble. At $z \sim 10$, only the heavy-seed super-Eddington model ({\sc hs-edd5}) and the PBH model are consistent with the observational lower limits from UHZ1 and GHZ9, providing a strong test of seeding and growth models at these early epochs. By $z \sim 5-7$, all of the models except Eddington limited light seeds ({\sc ls-edd1}) reproduce the observed bolometric luminosity functions within uncertainties. We caution bolometric luminosity corrections (used to convert Balmer line luminosities to a bolometric luminosity) are under revision, given the increasing wavelength ranges being probed for early black holes. Statistical samples of complete black hole SEDs will be crucial in determining the black hole mass/Balmer line-bolometric corrections at these early epochs.

\item The black hole mass scales with both the stellar mass and halo mass for all models at $z \sim 5-10$; the {\sc ls-edd5} and PBH models show the steepest and shallowest slopes of the black hole mass-stellar mass relation, respectively. The high black hole mass to stellar mass ratios ($\sim 0.3-1$) inferred from JWST at $z \sim 5-10$ pose a challenge for black hole models: light seeding models are unable to explain the presence of systems with $\mbh \sim M_*$ at $z \sim 10$, even in the super-Eddington case. While the heavy seeding Eddington-limited model can reproduce the bulk of the objects, it can not explain the existence of black holes with $\mbh \sim 10^7 \msun$ that are over-massive with respect to their hosts, with $M_* <10^7\msun$, at $z \sim 7-10$. Overall, only the {\sc hs-edd5} and PBH models are in accord with the data at {\rm all} $z \sim 5-10$. We also note the {\sc lm-edd1} and PBH models show the largest and smallest halo mass-to-black hole mass ratios, respectively.  

\item Metallicity scales with both black hole and stellar masses. Due to their lowest black hole to halo mass relations, galaxies in PBH models are the most affected by feedback and therefore show the lowest metallicity values for a given black hole mass, or stellar mass at $z \sim 5-7$. Indeed, only the PBH model is consistent with {\it all three axes} of the black hole mass-stellar mass-metallicity relation observed by \citet{tripodi2025} and \citet{maiolino2025}. However, it yields a slightly higher stellar mass than that inferred for the object observed by \citet{Napolitano2025_Xray} which is in better agreement with the {\sc hs-edd1} model. Globally, only the {\sc hs-edd1} and PBH models are in broad agreement with the metallicity relations at $z \sim 5-10$. 

\item PBH-seeded black holes inhabit systematically lower-mass halos than astrophysically-seeded ones of equivalent mass. The PBH model uniquely predicts a negative $\mbh/M_*$ trend with halo mass, in contrast to all astrophysical models where the value of $\mbh/M_*$ increases with halo mass. At $z \sim 7$, selecting systems with $\mbh/M_* > 0.1$ and $\lbol \sim 10^{44-46} \ergs$ that show a {\it negative} black hole to stellar mass ratio and reside in $10^{9-11}\msun$ halos can be used as possible indicators of PBH seeding models. 
\end{itemize}

There are a number of caveats that we now reiterate. The first concerns observationally-inferred quantities including the black hole mass, stellar mass, and metallicities, pinning down which urgently require statistical samples of black hole SEDs and host metallicities. In terms of theoretical assumptions, black holes in the first billion years probably arise from a mix of astrophysical (light and heavy) seeds, and cosmological seeds, complicating the interpretations presented here. We also caution that the mass spectrum of PBHs remains an open problem. We have also assumed super-Eddington accretion rates purely based on the available gas mass. As shown by a number of works, given that the Eddington limit assumes spherical accretion, super-Eddington accretion rates are possible, at least for extended periods, via geometrically thick disk accretion modes \citep{davis2020,safarzadeh2020} which needs to be explored further in semi-analytic models. In this work, we have also limited ourselves to non-spinning black holes; we aim to expand our calculations to include spinning systems in the future. We have assumed coupling between black hole feedback and all of the gas content of the halo, an assumption we hope to refine in future works. Finally, it is important to note that whilst the clustering of halos formed from primordial density perturbations is well known, the clustering of structures seeded by PBHs critically depends on their exact formation mechanism \citep[e.g.][]{matsubara2019, animali2024} and will likely be quite different from that in ``standard" structure formation.

Overall, the only model that can be conclusively ruled out as of now is one where all black holes are seeded by light seeds that can accrete at Eddington-limited rates. In general, the {\sc hs-edd5} and PBH models are most consistent with the lower limits on the bolometric luminosity function at $z \sim 10$, while the {\sc hs-edd1} and PBH models are in better agreement with the metallicities inferred for early systems. Future observations will be crucial in distinguishing between astrophysical and cosmological seeding models. These include: (i) an over-abundance of low-mass black holes $\sim 10^{3-4} \msun$ in the black hole mass function using gravitational lensing; (ii) extremely low metallicities for systems with high black hole to stellar mass ratios; (iii) negative black hole to stellar mass ratios as a function of the halo mass. Clustering signals via forthcoming {\it Roman} and {\it Euclid} observations will be crucial in obtaining hints on the halo mass; and (iv) deep X-ray observations with e.g. {\it NewAthena} to verify the black hole nature of these early systems.  
\begin{acknowledgements}
P. Dayal warmly acknowledges support from an NSERC discovery grant (RGPIN-2025-06182). She thanks M. Giavalisco and A. Mazumdar for illuminating discussions.

\end{acknowledgements}

\bibliographystyle{apsrev4-1}

\bibliography{bh}

\end{document}